# Experimental signatures of a three-dimensional quantum spin liquid in effective spin-1/2 $Ce_2Zr_2O_7$ pyrochlore

Bin Gao[1,11], Tong Chen[1,11], David W. Tam[1], Chien-Lung Huang[1], Kalyan Sasmal[2], Devashibhai T. Adroja[3], Feng Ye[4], Huibo Cao[4], Gabriele Sala[4], Matthew B. Stone[4], Christopher Baines[5], Joel A. T. Verezhak[5], Haoyu Hu[1], Jae-Ho Chung[1,6], Xianghan Xu[7], Sang-Wook Cheong[7], Manivannan Nallaiyan[8], Stefano Spagna[8], M. Brian Maple[2], Andriy H. Nevidomskyy[1], Emilia Morosan[1], Gang Chen[9,10] and Pengcheng Dai[1]\*

**A quantum spin liquid is a state of matter where unpaired electrons' spins, although entangled, do not show magnetic order even at the zero temperature. The realization of a quantum spin liquid is a long-sought goal in condensed-matter physics. Although neutron scattering experiments on the two-dimensional spin-1/2 kagome lattice $ZnCu_3(OD)_6Cl_2$ and triangular lattice $YbMgGaO_4$ have found evidence for the hallmark of a quantum spin liquid at very low temperature (a continuum of magnetic excitations), the presence of magnetic and non-magnetic site chemical disorder complicates the interpretation of the data. Recently, the three-dimensional $Ce^{3+}$ pyrochlore lattice $Ce_2Sn_2O_7$ has been suggested as a clean, effective spin-1/2 quantum spin liquid candidate, but evidence of a spin excitation continuum is still missing. Here, we use thermodynamic, muon spin relaxation and neutron scattering experiments on single crystals of $Ce_2Zr_2O_7$, a compound isostructural to $Ce_2Sn_2O_7$, to demonstrate the absence of magnetic ordering and the presence of a spin excitation continuum at 35 mK. With no evidence of oxygen deficiency and magnetic/non-magnetic ion disorder seen by neutron diffraction and diffuse scattering measurements, $Ce_2Zr_2O_7$ may be a three-dimensional pyrochlore lattice quantum spin liquid material with minimum magnetic and non-magnetic chemical disorder.**

A quantum spin liquid (QSL) state, where interacting quantum spins in a crystalline solid form a disordered state at zero temperature in much the same way as liquid water is in a disordered state, originates from Anderson's 1973 proposal that valence bonds between neighbouring spins in a two-dimensional (2D) triangular lattice can pair into singlets and resonate without forming long-range magnetic order[1]. As such a state may be important to the microscopic origin of high-transition-temperature superconductivity[2,3] and useful for quantum computation[4,5], the experimental realization of a QSL is a long-sought goal in condensed-matter physics. Although models supporting QSLs have been developed for 2D spin-1/2 kagome, triangular, honeycomb and 3D pyrochlore lattice systems[6–10], a common feature to all QSLs is the presence of deconfined spinons, the elementary excitations from the entangled ground state that carry spin $S=\frac{1}{2}$ and thus are fractionalized quasiparticles, fundamentally different from the $S=1$ spin waves in conventional 3D ordered magnets.

In 1D antiferromagnetic spin-1/2 chain compounds such as $KCuF_3$, the deconfined spinons have been unambiguously measured as a spin excitation continuum by inelastic neutron scattering experiments[11]. In 2D spin-1/2 triangular organic salts such as $\kappa$-$(ET)_2Cu_2(CN)_3$ (ref. [12]) and $EtMe_3Sb[Pd(dmit)_2]_2$ (ref. [13]), while nuclear magnetic resonance measurements indicate the presence of a QSL, there have been no inelastic neutron scattering experiments to search for a spin excitation continuum due to the lack of large single crystals. While continua of spin excitations are seen by inelastic neutron scattering in the 2D spin-1/2 kagome lattice $ZnCu_3(OD)_6Cl_2$ (refs. [14,15]) and in an effective spin-1/2 triangular lattice magnet $YbMgGaO_4$ (refs. [16,17]), the magnetic/non-magnetic site disorder in the kagome lattice[18] and non-magnetic site disorder in the triangular lattice[19] case complicate the interpretation of the data[15,20–24]. Recently, the Heisenberg quantum magnet $Ca_{10}Cr_7O_{28}$, where the spin-1/2 $Cr^{5+}$ ions form a distorted kagome bilayer structure, revealed clear evidence for a 3D QSL[25]. There are also signatures of a 3D QSL in the hyperkagome lattice compound $PdCuTe_2O_6$ (ref. [26]) and the spin-1 antiferromagnet $NaCaNi_2F_7$ (ref. [27]). Nevertheless, there is no consensus on the experimental confirmation of a QSL with spin quantum number fractionalization in a 3D pyrochlore lattice spin-1/2 magnet.

In 3D rare-earth pyrochlores such as $Ho_2Ti_2O_7$, Ising-like magnetic moments decorate a lattice of corner-sharing tetrahedra (Fig. 1a) and form the '2-in/2-out' spin ice arrangement, analogous to the '2-near/2-far' rule of the covalent $2H^+$–$O^{2-}$ bonding distances in water ice, to stabilize classical spin liquids[28,29]. A key feature of classical spin ice systems is the presence of low-temperature residual magnetic entropy (equivalent to a ground-state entropy per spin of $1/2\ln(3/2)$) analogous to the Pauling estimate of the residual entropy for water ice[29,30]. Due to the extensive ground-state degeneracy, the magnetic entropy $S_{mag}$ of a classical spin ice without a magnetic field does not saturate to $R\ln 2$ for an effective spin-1/2 system and is instead set by $S_{mag}=R(\ln 2 - 1/2\ln(3/2))$ in the high-temperature limit, where $R$ is the ideal gas constant[29,30]. In the presence of quantum fluctuations, a QSL state could emerge in the so-called quantum spin ice regime characterized by the emergent $U(1)$ quantum electrodynamics[30]. Here, the QSL state has a $U(1)$ gauge degree of freedom, similar to the gauge symmetry of Maxwell's equations, and the emergent photon-like gapless excitations[31]. Up to now, most works have considered the degenerate spin ice manifold in the classical limit as the starting point for realizing the $U(1)$ QSL on introducing quantum fluctuation. However, within a mean-field theory, the $U(1)$ QSL could extend much beyond the ice limit and thus does not necessarily produce any phenomena related to classical spin ice in the finite-temperature regime. Thus, the candidate pyrochlore QSL materials that do not show classical spin ice characteristics such as the Pauling entropy at finite temperatures may still be a $U(1)$ QSL or other QSL not far from it[32].

[1]Department of Physics and Astronomy, Rice University, Houston, TX, USA. [2]Department of Physics, University of California, San Diego, San Diego, CA, USA. [3]ISIS Facility, STFC Rutherford-Appleton Laboratory, Didcot, UK. [4]Neutron Scattering Division, Oak Ridge National Laboratory, Oak Ridge, TN, USA. [5]Laboratory for Muon-Spin Spectroscopy, Paul Scherrer Institut, Villigen PSI, Switzerland. [6]Department of Physics, Korea University, Seoul, Korea. [7]Rutgers Center for Emergent Materials and Department of Physics and Astronomy, Rutgers University, Piscataway, NJ, USA. [8]Quantum Design Inc., San Diego, CA, USA. [9]Department of Physics and Center of Theoretical and Computational Physics, The University of Hong Kong, Hong Kong, China. [10]State Key Laboratory of Surface Physics and Department of Physics, Fudan University, Shanghai, China. [11]These authors contributed equally: Bin Gao, Tong Chen.
\*e-mail: pdai@rice.edu

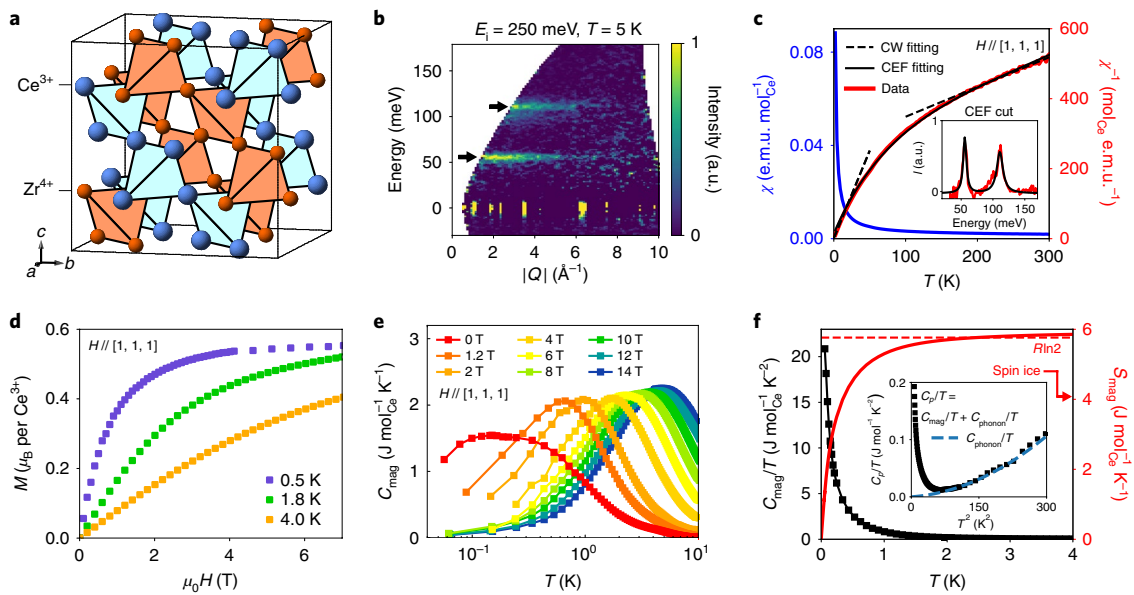

**Fig. 1 | Summary of crystal structure, CEF levels, d.c. susceptibility and specific heat of $Ce_2Zr_2O_7$. a**, A schematic of the structure of $Ce_2Zr_2O_7$ with lattice parameters $a=b=c=10.71$ Å. The blue ions are magnetic $Ce^{3+}$ (A site) and the red ions are non-magnetic $Zr^{4+}$ (B site). **b**, Inelastic neutron scattering from powder samples of $Ce_2Zr_2O_7$ with an incident neutron beam energy of $E_i=250$ meV at 5 K. Phonon contributions and other background scattering measured in the non-magnetic analogue $La_2Zr_2O_7$ were subtracted. Two Kramers doublets at 55.8 and 110.8 meV are clearly seen. **c**, The d.c. magnetic susceptibility $\chi$ (blue) and $1/\chi$ (red). The applied field is 1 kOe along the [1, 1, 1] direction. The black dashed lines are the low–high temperature Curie–Weiss (CW) fit to the data. The red curve in the inset shows the energy dependence of the CEF levels for integrated wavevectors from 4 to 6 Å$^{-1}$. The solid black lines in the inset and main panel are CEF fits to the data. **d**, Magnetization ($M$) as a function of applied magnetic field ($H$) in tesla along the [1, 1, 1] direction at different temperatures. **e**, The magnetic contribution to the specific heat $C_{mag}$ for different magnetic fields along the [1, 1, 1] direction. The lattice contribution $C_{phonon}$ has been subtracted. **f**, $C_{mag}/T$ versus $T$ plot and the magnetic entropy $S_{mag}$ calculated from $C_{mag}/T$ in zero field. $S_{mag}$ reaches the entropy for a two-level system ($R\ln 2$) instead of the spin ice entropy $(1-(1/2)(\ln(3/2)/\ln 2)\,R\ln 2$. The inset shows a $C/T$ versus $T^2$ plot. The data between 10 and 20 K are fitted with $C_{phonon}/T=\beta T^2+\alpha T^4$.

Recently, the Ce-based pyrochlore stannate $Ce_2Sn_2O_7$ has been proposed as a 3D QSL from thermodynamic and muon spin relaxation (μSR) measurements on powder samples[33]. The Ce local moment in $Ce_2Sn_2O_7$ is the peculiar dipole–octupole doublet that may support distinct symmetry-enriched $U(1)$ QSLs[34,35]. However, in the absence of single crystals of $Ce_2Sn_2O_7$, there have been no inelastic neutron scattering experiments to search for the expected spin excitation continua. To overcome this problem, we used the floating-zone method to grow high-quality single crystals of $Ce_2Zr_2O_7$ (see Methods and Supplementary Fig. 1), an isoelectronic/isostructural compound of $Ce_2Sn_2O_7$ (ref. [36]). In the stoichiometric $Ce_2Zr_2O_7$ pyrochlore structure with the $Fd\bar{3}m$ space group, cerium ions stabilize in the magnetic $Ce^{3+}$ ($4f^1, {}^2F_{5/2}$) state in the crystal field of eight oxygen anions (Supplementary Fig. 2a). $Ce^{3+}$ with $J=5/2$ has an odd number of $f$ electrons and the crystal electric field (CEF) potential from oxygen will split them into three Kramers doublets[10]. Figure 1b shows the inelastic neutron scattering spectra from the $Ce^{3+}$ CEF levels (see also Supplementary Fig. 5), revealing two excited states at ~55 and ~110 meV. Based on the point-group symmetry at the $Ce^{3+}$ atomic site and using the Stevens operator formalism (see Methods), the CEF Hamiltonian with the quantization axis along the local [1, 1, 1] direction can be written as $H_{CEF}=B_2^0\hat{O}_2^0+B_4^0\hat{O}_4^0+B_4^3\hat{O}_4^3+B_6^0\hat{O}_6^0+B_6^3\hat{O}_6^3+B_6^6\hat{O}_6^6$, where $B_2^0$, $B_4^0$ and $B_6^0$ are the second-, fourth- and sixth-order CEF parameters, and $\hat{O}_2^0$, $\hat{O}_4^0$ and $\hat{O}_6^0$ are the corresponding Stevens operator equivalents, respectively[10]. Since $Ce^{3+}$ has $J=5/2$ for the ground-state multiplet, the maximum allowed terms in the CEF Hamiltonian are less than $2J$, meaning that the sixth-order terms are zero: $B_6^0=B_6^3=B_6^6=0$. Using the CEF Hamiltonian to fit the two inelastic excitations in Fig. 1b, we find $B_2^0=-1.27$, $B_4^0=0.32$ and $B_4^3=-1.86$ meV (see inset of Fig. 1c), with the $Ce^{3+}$ ground-state doublet being $J_z=\pm 3/2$, where $J_z$ is along the [1, 1, 1] direction (Fig. 1a). As each state in the doublet is a 1D irreducible representation of the $D_{3d}$ point group, the $Ce^{3+}$ ground-state doublet in $Ce_2Zr_2O_7$ is the dipole–octupole doublet, identical to that of $Ce_2Sn_2O_7$ (ref. [34]). The degeneracy of the $Ce^{3+}$ ground-state doublet here is protected by time-reversal symmetry. The $Ce^{3+}$ dipole–octupole doublet is very different from the Kramers doublet of the $Yb^{3+}$ ground state in $Yb_2Ti_2O_7$ and the non-Kramers doublet of the $Pr^{3+}$ ground state in $Pr_2Zr_2O_7$, where the doublet in both cases forms a 2D irreducible representation of the $D_{3d}$ point group[7,10]. Instead, the Ce doublet here is analogous to the ground-state doublet of $Nd^{3+}$ in $Nd_2Zr_2O_7$ (ref. [37]).

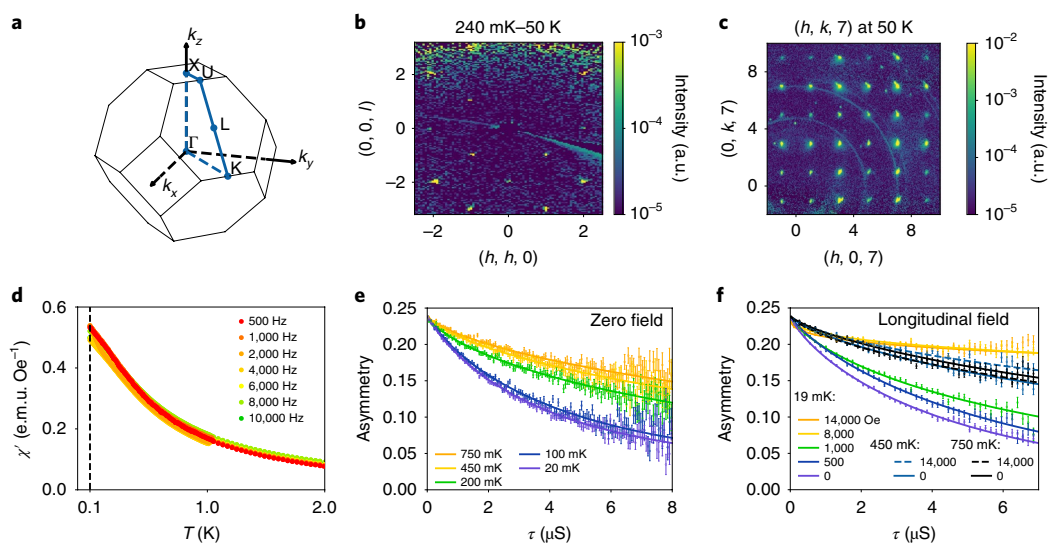

**Fig. 2 | Brillouin zone, diffuse neutron scattering, a.c. susceptibility and µSR data of $Ce_2Zr_2O_7$. a**, A schematic diagram of the Brillouin zone of the pyrochlore lattice and the $(h, h, l)$ zone (blue). **b**, The wavevector dependence of the diffuse neutron scattering between 240 mK and 50 K in the $(h, h, l)$ zone. Spin ice pyrochlores show clear diffuse scattering[41–44]. **c**, The diffuse neutron scattering in the $(h, k, 7)$ plane of a $Ce_2Zr_2O_7$ single crystal at 50 K. There is no evidence of diffuse scattering induced by oxygen vacancy defects as seen in $Y_2Ti_2O_{7-x}$ (ref. [38]). **d**, The a.c. magnetic susceptibility $\chi'$ as a function of frequency. There is no evidence of spin freezing behaviour above 100 mK. **e,f**, Zero-field and longitudinal field data in the usual histogram form $A(t)$; the solid lines are fits to the function $A(t) = A_0 \exp\left[-\left(\frac{t}{T_1}\right)^\beta\right] + A_{bg}$. The error bars represent one standard deviation.

The blue solid line in Fig. 1c shows the temperature dependence of the magnetic susceptibility $\chi(T)$ with no evidence of a magnetic transition above 0.5 K. The red and black solid lines are experimental and calculated $1/\chi(T)$ using the fitted CEF parameters from Fig. 1b, respectively. The dashed lines are low- and high-temperature Curie–Weiss fits to the data, giving a Curie–Weiss temperature of $\theta_{CW} = -0.57 \pm 0.01$ K, a Curie constant of $\sim 0.2$ emu mol$^{-1}$ K$^{-1}$ and an effective moment of $\sim 1.28\mu_B$ per $Ce^{3+}$ from the low-temperature fit. Therefore, the ground state of $Ce_2Zr_2O_7$ is an effective spin-1/2 Kramers doublet. Due to the hybrid multipolar nature, the $Ce^{3+}$ local moment has Ising-like anisotropic g-tensors with a parallel component (along the $[1, 1, 1]$ direction) $g_\parallel = 2.57$ and a perpendicular component $g_\perp = 0$, different from $Er_2Ti_2O_7$ and $Yb_2Ti_2O_7$ XY pyrochlores where all three components of the effective spin carry dipole moments[10].

Figure 1d shows isothermal magnetization curves, $M(H)$, for $Ce_2Zr_2O_7$ at 4 K, 1.8 K and 0.5 K; the isothermal magnetization saturates at approximately half of the value of the effective magnetic moment observed in the moderate temperature plateau similar to that of $Ce_2Sn_2O_7$ (Supplementary Fig. 2b)[33]. Figure 1e shows the temperature dependence of the magnetic contributions to the specific heat as a function of applied magnetic fields along the $[1, 1, 1]$ direction. At zero field, we see a broad peak centred around 0.2 K, which may arise from the onset of coherent quantum fluctuations similar to the case of the 3D QSL $Ca_{10}Cr_7O_{28}$ (ref. [25]), different from the XY pyrochlores with long-range magnetic order[10]. After subtracting the phonon contributions to the specific heat and assuming a power-law extrapolation to 0 K below the lowest measured temperature of 50 mK, we calculate the magnetic entropy by integrating the magnetic contribution to the specific heat $C_{mag}/T$ from $T = 0$ to 4 K (inset, Fig. 1f). The temperature dependence of the magnetic entropy saturates at $S_{mag} \approx 1.01R\ln 2$ at 4 K, consistent with a free-spin system (Fig. 1f). The absence of an entropy plateau and Pauling entropy in the magnetic entropy curve indicates a key difference of $Ce_2Zr_2O_7$ from a classical spin ice (Fig. 1f). On application of a magnetic field, the broad peak evolves into a Schottky-type anomaly and shifts upwards in temperature consistent with the field-induced splitting of the effective spin-1/2 doublet. Therefore, $Ce_2Zr_2O_7$ is probably a non-spin-ice-type pyrochlore QSL.

To demonstrate that our single crystals of $Ce_2Zr_2O_7$ have a stoichiometric pyrochlore structure with $Ce^{3+}$ ions and no magnetic order approaching zero temperature, we carried out single-crystal X-ray and neutron diffraction experiments by measuring more than 700 and 100 Bragg peaks, respectively (see Methods). The fitting outcome reveals anti-site disorder between Ce and Zr of 4(1)% (Fig. 1a) and oxygen occupancy of 98(3)%; both are within the errors of the measurements (see Supplementary Table 1). We have further carried out single-crystal neutron diffraction and diffuse scattering measurements within the $[h,h,l]$ scattering plane (Fig. 2a). Figure 2b shows a temperature difference map between 240 mK and 50 K, revealing no evidence of antiferromagnetic order down to 240 mK. In previous work on spin ice pyrochlores[29], it was found that oxygen vacancies in the pyrochlore structure can induce magnetic impurities and transform the nature of the magnetism[38]. In particular, neutron scattering measurements on single crystals of oxygen-deficient $Y_2Ti_2O_{7-x}$ reveal clear evidence for one of the oxygen site (O2) vacancy-induced diffuse scattering in the $(h, 0, 7) \times (0, k, 7)$ scattering plane[38] (see Supplementary Fig. 2a for location of the O2 site). We find no evidence for similar oxygen vacancy-induced diffuse scattering in $Ce_2Zr_2O_7$ (Fig. 2c and Supplementary Fig. 6).

In the triangular lattice magnet $YbMgGaO_4$(refs. [16,17]), the a.c. magnetic susceptibility shows a clear frequency dependence, which was interpreted as evidence for a spin glass freezing instead of a QSL[20]. To test whether $Ce_2Zr_2O_7$ may also have a spin glass ground state, we carried out measurements of the frequency dependence of the a.c. susceptibility measurements. As can be seen from Fig. 2d, the a.c. susceptibility increases with decreasing temperature, but shows no evidence of a phase transition above 100 mK and no frequency dependence from 500 Hz to 10000 Hz. Therefore, we conclude reliably that Ce$_2$Zr$_2$O$_7$ does not have a spin glass or spin ice ground state.

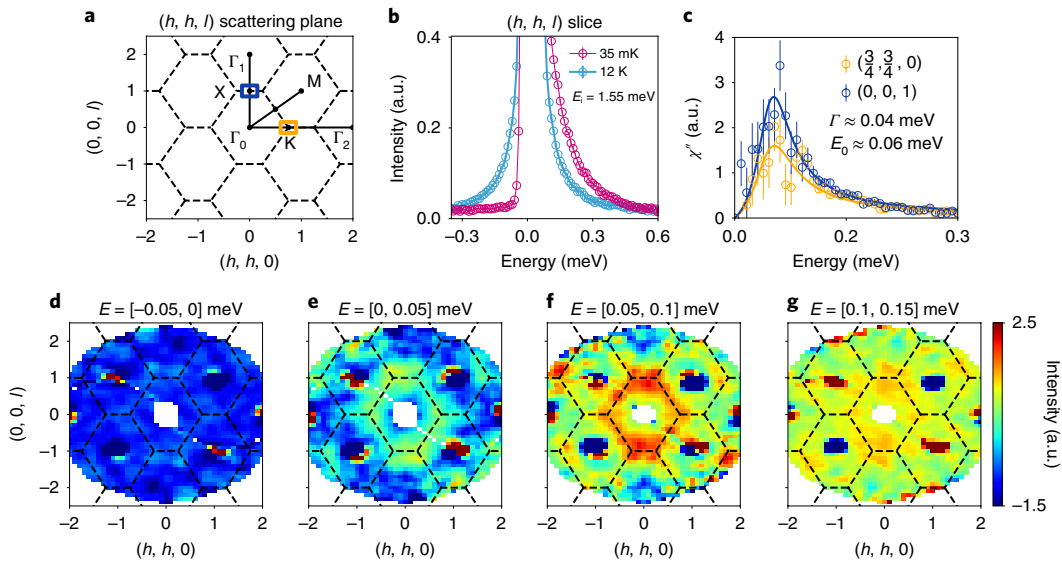

**Fig. 3 | Summary of energy and wavevector dependence of the spin excitations in Ce$_2$Zr$_2$O$_7$. a**, A schematic diagram of the (h, h, l) zone and high-symmetry directions. The dashed lines mark the zone boundary. The orange and blue boxes mark the wavevector integration range for magnetic scattering in **c**. **b**, Wavevector-averaged spectra within the (h, h, l) scattering plane at 35 mK and 12 K. The incident neutron energy is $E_i = 1.55$ meV. The wavevector integration ranges along the (h, h, 0) and (0, 0, l) directions are from −3.0 to 3.0 for both directions with a thickness of −0.05 to 0.05, all in reciprocal lattice units. **c**, The energy dependence of $\chi''$ obtained by subtracting 12 K scattering from 35 mK scattering. The data are obtained with $E_i = 3.32$ meV and the integration areas are shown in **a**. **d–g**, Wavevector dependence of the magnetic scattering at 35 mK and different energies obtained by using $E_i = 3.32$ meV. The magnetic excitations are seen in a large region of the Brillouin zone. With increasing energy, the magnetic excitations become increasingly diffusive. The data in **d–g** and Fig. 4a were collected from 0 to 180°, but expanded to 180 to 360°. No folding operations were performed to improve the statistics. The error bars in **b** and **c** represent one standard deviation.

Although our specific heat (Fig. 1e), magnetic susceptibility (Fig. 2d) and neutron diffraction (Fig. 2b) experiments on Ce$_2$Zr$_2$O$_7$ revealed no static magnetic or spin glass order above 100 mK, it is still important to test whether the system can order magnetically at lower temperatures using μSR experiments. Figure 2e shows the zero-field data at several temperatures from 20 mK to 750 mK. At all measured temperatures, we find no evidence for long-range magnetic order, which would have presented itself as oscillations or a fast-relaxing component. Instead, the data can be well described by a stretched exponential relaxation plus a background, similar to that of Ce$_2$Sn$_2$O$_7$ (ref. [33]). The temperature dependence of the relaxation rate shows a kink below 200 mK (Supplementary Fig. 4), suggesting slowing down of the spin dynamics below this temperature (see Methods). To determine whether the spin relaxation arises from a static internal field (due to static magnetic order) and is related to the magnetic field as suggested from the heat capacity data (Fig. 1e), we carried out longitudinal field measurements up to 1.4 T at different temperatures (Fig. 2f). If the relaxation is from a static internal field, it should be decoupled by application of about 100 G field in the longitudinal field measurements. Instead, our longitudinal field measurements show that the μSR rate at 20 mK is dominated by internal magnetic fluctuations, which is suppressed only at a high field above ~3,000 G (Fig. 2f), indicating that the observed relaxation arises from dynamic spin fluctuations. In addition, application of a magnetic field pushes the spin fluctuations to higher energies as seen in the field-dependent specific heat, resulting in a slower relaxation of the muon polarization (Fig. 2f).

Having established that Ce$_2$Zr$_2$O$_7$, similar to Ce$_2$Sn$_2$O$_7$ (ref. [33]), is a stoichiometric pyrochlore with an effective spin-1/2 Kramers doublet ground state and no static magnetic order above 20 mK, it will be important to determine the energy, wavevector and temperature dependence of the spin excitations in the system. For this purpose, we align the crystal in the (h, h, 0) × (0, 0, l) scattering plane (Figs. 2a and 3a). Figure 3b shows the energy dependence of the integrated neutron scattering at 35 mK and 12 K. On cooling from 12 K to 35 mK, the scattering intensity at 35 mK increases around the neutron energy loss side (red circles in Fig. 3b). Assuming that the temperature difference scattering between 35 mK and 12 K is entirely magnetic in origin, the imaginary part of the generalized dynamic spin susceptibility $\chi''(E)$ can then be calculated via $\chi''(E) = \left[1 - \exp\left(-\frac{E}{k_B T}\right)\right] S(E)$, where $S(E)$ is the magnetic scattering, $E$ is the neutron energy transfer and $k_B$ is Boltzmann's constant. Figure 3c shows the energy dependence of $\chi''$ near the K and X points in reciprocal space, which can be well described by $\chi''(E) \propto \Gamma E / [(E - E_0)^2 + \Gamma^2]$ with $\Gamma = 0.04$ mV and $E_0 = 0.06$ meV characterizing the damping and energy scale of the spin excitations, respectively. Figure 3d–g summarizes the wavevector dependence of spin excitations in the (h, h, l) zone at 35 mK for different spin excitation energies. As expected, we see no evidence of magnetic scattering for energy transfers from −0.05 to 0 meV (Fig. 3d). On moving to $E = 0.025 \pm 0.025$ meV, we see a continuum of spin excitations around the Brillouin zone boundary of the system (Fig. 3e). On further increasing the energy to $E = 0.075 \pm 0.025$ meV, the spin excitation continuum remains the same shape but with increased intensity (Fig. 3f). Finally, spin excitations at $E = 0.125 \pm 0.025$ meV are less clear but may still have the same wavevector dependence as those at other energies. In the temperature range of the Pauling entropy plateau for classical and quantum spin ice pyrochlore lattices, one would expect to find pinch points around M, $\Gamma_1$ and $\Gamma_2$ points in reciprocal space (Fig. 3a)[31]. However, we find no clear evidence of such pinch points in our data (Fig. 3d–g).

Figure 4a shows the wavevector dependence of the energy-integrated magnetic scattering from $E = −0.05$ to 0.15 meV. Consistent with the data in Fig. 3d–g, the magnetic scattering reveals a broad continuum confined to the zone boundary. Plotted in Fig. 4b–d are wavevector–energy dispersions of the spin excitations and wavevector cuts at different energies. Along all directions, the magnetic scattering is confined to below 0.2 meV, an energy scale considerably smaller than that of the spin excitations reported in classical spin ice and quantum XY pyrochlores[10,29].

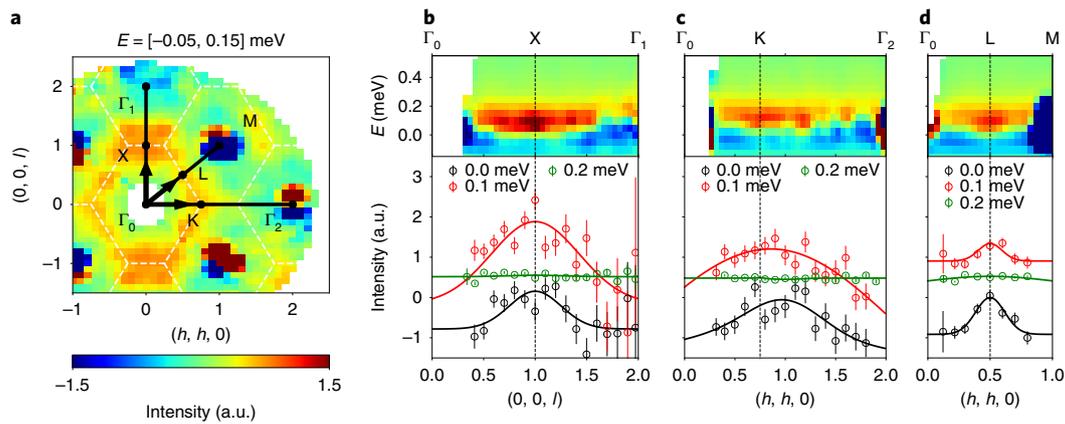

**Fig. 4 | The energy and wavevector dispersion of magnetic excitations in Ce$_2$Zr$_2$O$_7$. a**, The wavevector dependence of the integrated magnetic scattering from −0.05 to 0.15 meV at 35 mK. **b–d**, The dispersions of the spin excitations along different high-symmetry directions and constant-energy cuts along the (0, 0, $l$), ($h$, $h$, 0) and ($h$, $h$, $h$) directions. The black, red and green symbols mark cuts at energies of $E = 0 \pm 0.05$, $0.1 \pm 0.05$ and $0.2 \pm 0.05$ meV, respectively. The solid lines are guides to the eye and the error bars represent one standard deviation.

Assuming that the Ce$^{3+}$ ions are dipole–octupole doublets, the generic model for the Ce-based pyrochlore lattice is the XYZ spin model with the Hamiltonian $H = \sum_{ij} J_z S_i^z S_j^z + J_x S_i^x S_j^x + J_y S_i^y S_j^y + J_{xz}(S_i^x S_j^z + S_i^z S_j^x)$, where $S^z$ is the magnetic dipole moment defined along the [1, 1, 1] direction, and $S^x$ and $S^y$ are the magnetic octupole moments[34,35]. Unlike the spatially anisotropic model for Yb$_2$Ti$_2$O$_7$ and Er$_2$Ti$_2$O$_7$ (refs. [39,40]), the XYZ model here is uniform on every bond while the interaction anisotropy occurs in the effective spin space and arises from the dipole–octupole nature of the local moments. As far as the lattice symmetry is concerned, $S^x$ transforms identically as the dipole moment $S^z$, and is thus regarded as the dipole moment in certain cases. Here, we regard $S^x$ and $S^y$ as the octupole moments. As only the dipole moment $S^z$ couples linearly with the magnetic field, the Curie–Weiss temperature merely measures the $J_z$ coupling and therefore $J_z = 2\theta_{CW} = -1.14 \pm 0.02$ K. The remaining three couplings cannot be simply determined from the current measurements. One potential approach is to apply magnetic fields along various high-symmetry directions to polarize $S^z$ and induce magnetism. Even though only the $S^z$–$S^z$ spin correlations are detectable by inelastic neutron scattering, the field-induced $S^z$ component actually polarizes the $S^x$ component via the crossing coupling $J_{xz}$, and thus inelastic neutron scattering experiments in the field-induced ordered state contain both the coherent spin-wave excitations and two-magnon continuum from which one may extract the exchange coupling constants.

Theoretically, the XYZ model for the dipole–octupole doublets has two symmetry-enriched $U(1)$ QSLs dubbed a dipolar QSL and an octupolar QSL[34,35]. They are distinct by the roles of the dipole and octupole components in each phase. Application of an external magnetic field can in principle induce an emergent Anderson–Higgs transition by condensing the spinons in the octupolar QSL[35]. Although the present data cannot distinguish which QSL is actually realized in Ce$_2$Zr$_2$O$_7$, the absence of Pauling entropy from heat capacity data (Fig. 1f) and the absence of spin freezing in a.c. susceptibility measurements (Fig. 2d) indicate that the QSL in Ce$_2$Zr$_2$O$_7$ is not in the spin ice regime. In the un-frustrated regime of the XYZ model, where there is no sign problem for quantum Monte Carlo simulation, the system should be either in the $U(1)$ QSL phase of the spin ice regime or in an ordered phase[34,35]. Since the QSL in Ce$_2$Zr$_2$O$_7$ is not in the spin ice regime, the relevant XYZ model should be in the frustrated regime where QSLs are more robust[32]. Since the XYZ model brings interaction to the spinons in the $U(1)$ QSL phase, spinon pairing in the un-frustrated regime could stabilize a fully gapped $Z_2$ QSL[32], and such pairing in the frustrated regime may form a spinon 'superconductor'.

To summarize, we have discovered a 3D pyrochlore lattice QSL candidate, Ce$_2$Zr$_2$O$_7$, that has all of the hallmarks of a QSL but with considerably reduced complications of magnetic and lattice disorder, or oxygen vacancy, and up to 4% anti-site disorder[38]. We further demonstrate that Ce$_2$Zr$_2$O$_7$ may be a non-spin-ice pyrochlore QSL and differ from the prevailing examples of spin-ice-based pyrochlore QSL candidate materials[10,41–44]. Our measurements indicate the absence of long-range order or spin glass behaviour, while the neutron spectroscopy finds conclusive evidence for a continuum of fractionalized excitations, as theoretically expected in a QSL.

*Note added in proof*: After submission of the present work, we became aware of a related neutron scattering work on Ce$_2$Zr$_2$O$_7$ (ref. [45]).

### Online content

Any methods, additional references, Nature Research reporting summaries, source data, statements of code and data availability and associated accession codes are available at https://doi.org/10.1038/s41567-019-0577-6.

## Acknowledgements
We thank Y. Su, Y. J. Uemura, J. A. Rodriguez and Z. Ma for helpful discussions. The neutron scattering work at Rice is supported by US DOE BES DE-SC0012311 (P.D.). The single-crystal-growth work at Rice is supported by the Robert A. Welch. Foundation under grant no. C-1839 (P.D.). E.M. and C.-L.H. acknowledge the support from US DOE BES DE-SC0019503. A.H.N. acknowledges the support of the Robert A. Welch Foundation grant no. C-1818 and NSF CAREER grant no. DMR-1350237. Research at UCSD was supported by the US DOE BES DE-FG02-04ER46105 (M.B.M.). The work of J.-H.C. was supported by the National Research Foundation of Korea (NRF-2017K1A3A7A09016303). This research used resources at the Spallation Neutron Source and the High Flux Isotope Reactor, DOE Office of Science User Facilities operated by the ORNL. H.C. acknowledges the support of US DOE BES Early Career Award KC0402010 under contract DE-AC05-00OR22725. Crystal growth by B.G., X.X. and S.-W.C. at Rutgers was supported by the visitor programme at the Center for Quantum Materials Synthesis, funded by the Gordon and Betty Moore Foundation's EPiQS initiative through grant GBMF6402, and by Rutgers University. G.C. acknowledges the support from the Ministry of Science and Technology of China with grant no. 2016YFA0301001 and 2016YFA0300500.

## Author contributions
P.D., B.G. and G.C. conceived the project. B.G., T.C., X.X. and S.-W.C. prepared the samples. d.c. magnetic susceptibility and heat capacity measurements were performed by C.-L.H. in the laboratory of E.M. a.c. susceptibility measurements were carried out by K.S., M.N., S.S. and M.B.M. μSR measurements were carried out and analysed by D.W.T. with help from C.B. and J.A.T.V. The CEF level measurements and analysis were performed by B.G., M.B.S. and D.T.A. X-ray and neutron diffraction experiments were carried out and analysed by H.C. Neutron diffuse scattering experiments were carried out by B.G. and F.Y. Inelastic neutron scattering experiments were carried out and analysed by B.G., T.C., G.S., H.H. and J.-H.C. Theoretical analysis was performed by G.C. A.H.N. helped with the initial theoretical analysis. The entire project was supervised by P.D. The manuscript was written by P.D., B.G., T.C., A.H.N. and G.C. All authors made comments.

## Competing interests
The authors declare no competing interests.

## Additional information
**Supplementary information** is available for this paper at https://doi.org/10.1038/s41567-019-0577-6.

**Reprints and permissions information** is available at www.nature.com/reprints.

**Correspondence and requests for materials** should be addressed to P.D.

**Peer review information:** *Nature Physics* thanks Kazushi Kanoda and the other, anonymous, reviewer(s) for their contribution to the peer review of this work.

**Publisher's note:** Springer Nature remains neutral with regard to jurisdictional claims in published maps and institutional affiliations.

## Methods

**Sample growth.** Polycrystalline $Ce_2Zr_2O_7$ was synthesized using a solid-state reaction method. Stoichiometric powders of $CeO_2$ (Alfa Aesar, 99.9%) and ZrN (Alfa Aesar, 99.9%) were mixed, ground, pelletized and sintered at 1,400 °C in air. The sintered pellet was then ground, pelletized and sintered in a forming gas (8% $H_2$ in Ar) flow at 1,400 °C for 20 h. Single crystals of $Ce_2Zr_2O_7$ were synthesized using a laser diode floating-zone furnace at the Center for Quantum Materials Synthesis of Rutgers University (Supplementary Fig. 1a). Polycrystalline $La_2Zr_2O_7$ was also synthesized using a solid-state reaction method. Stoichiometric powders of $La_2O_3$ (Alfa Aesar, 99.9%) and $ZrO_2$ (Alfa Aesar, 99.9%) were mixed, pelletized and sintered in air at 1,400 °C and 1,600 °C for the first and second sintering steps, respectively. The X-ray diffraction pattern of the ground $Ce_2Zr_2O_7$ single crystal revealed its pure pyrochlore phase (Supplementary Fig. 1e), with a lattice constant of $a = 10.70$ Å and the $Fd\bar{3}m$ space group. The Laue patterns of the single-crystalline $Ce_2Zr_2O_7$ in the [1,0,0] and [1,1,1] directions (Supplementary Fig. 1c,d) are sharp and clear, revealing the good quality of the crystals.

**X-ray and neutron diffraction experimental set-ups.** X-ray single-crystal diffraction experiments were carried out at Oak Ridge National Laboratory (ORNL); the measured crystal was carefully suspended in Paratone oil and mounted on a plastic loop attached to a copper pin/goniometer. The reported single-crystal X-ray diffraction data were collected with molybdenum $K_\alpha$ radiation ($\lambda = 0.71073$ Å) using a Rigaku XtaLAB PRO diffractometer equipped with a Dectris Pilatus 200 K detector and an Oxford N-HeliX cryocooler. More than 700 diffraction peaks were collected and refined using Rietveld analysis. The outcome is shown in Supplementary Table 1. Single-crystal neutron diffraction measurements were performed at the HB-3A DEMAND single-crystal neutron diffractometer at the High Flux Isotope Reactor, ORNL. The neutron wavelength of 1.550 Å was used from a bent perfect Si-220 monochromator[46,47]. The structure refinements were carried out with the FullProf Suite[48]. By carrying out both X-ray and neutron diffraction experiments, we were able to accurately determine the stoichiometry of Ce, Zr and O as shown in Supplementary Table 1.

**Bond valence sum and saturation value of magnetization.** Using crystallographic information obtained from X-ray and neutron diffraction experiments, we can estimate the valence of Ce using the bond valence sum method, where the valence $V$ of an atom is the sum of the individual bond valences $v_i$ surrounding the atom via $V = \sum v_i$. The individual bond valences in turn are calculated from the observed bond lengths via $v_i = \exp\left(\frac{R_0 - R_i}{b}\right)$, where $R_i$ is the observed bond length, $R_0$ is a tabulated parameter expressing the (ideal) bond length when the element $i$ has exactly valence 1, and $b$ is an empirical constant, typically 0.37 Å. In the case of $Ce_2Zr_2O_7$, each Ce atom is surrounded by eight oxygen atoms as shown in Supplementary Fig. 2a. The eight oxygen atoms have two different sites (see the positions of O1 and O2 in Supplementary Table 1). The two O2 sites are at the centre of the Ce tetrahedra, denoted by light blue spheres in Supplementary Fig. 2a, while the other six O1 sites are denoted by brown spheres. According to the single-crystal X-ray diffraction refinement results, the bond length of the two Ce–O2 bonds is 2.313 Å, and the bond length of the other six Ce–O1 bonds is 2.576 Å, as seen in Supplementary Fig. 2a. For a $Ce^{3+}$ cation surrounded by $O^{2-}$, $R_0$ equals 2.121 Å[4]. This gives $V = 2.945$, very close to 3, thus confirming that Ce indeed has a 3+ ionic state.

Figure 1d shows the saturation value of the magnetization, which is about $0.5\mu_B$ per $Ce^{3+}$, less than that expected for an effective spin-1/2 system. The lower value of the magnetization is due to the fact that field-induced $Ce^{3+}$ moments are aligned along the local [1,1,1] direction[49,50], as shown in Supplementary Fig. 2b. With this consideration, the total saturation moment for a $Ce^{3+}$ tetrahedron is consistent with a spin-1/2 moment.

**Magnetic susceptibility measurements.** Low-temperature a.c. magnetic susceptibility measurements were performed using a PPMS DynaCool dilution refrigerator equipped with the a.c. susceptibility measurement option located at Quantum Design. An a.c. field of 0.2 Oe was applied along the [1,1,1] direction of the $Ce_2Zr_2O_7$ single crystal. The frequency ranged from 100 Hz to 10 kHz and the temperature was down to 100 mK. The smooth monotonic increase of the a.c. susceptibility with decreasing temperature suggests that there is no long-range magnetic order or spin glass order in this temperature range (above 100 mK). Magnetic or spin glass ordering should give a peak in the a.c. magnetic susceptibility, which could, in principle, occur at a lower temperature. We note that our μSR experiments indicate no spin glass or magnetic order above 19 mK.

**Heat capacity measurements.** The specific heat was measured down to 0.05 K using a thermal-relaxation method in a DynaCool (Quantum Design) with the magnetic field applied along the [1,1,1] direction of the $Ce_2Zr_2O_7$ single crystal at Rice University. The total specific heat of $Ce_2Zr_2O_7$ is expressed as a sum of magnetic and lattice contributions: $C_p = C_{mag} + C_{lat}$. We fit the specific heat of $Ce_2Zr_2O_7$ between $T = 10$ and 20 K with $C_{lat} = \beta T^3 + \alpha T^5$. We also measured $La_2Zr_2O_7$ as the background and found no evidence of a magnetic/electronic contribution below 2 K.

The broad peak of $C_{mag}$ in zero field of $Ce_2Zr_2O_7$ may be a ubiquitous signature of the QSL ground state, signalling the onset of quantum coherent fluctuations as discussed in ref. [26]. Similar features of $C_{mag}$, both in zero field and in field, have been seen in other spin liquid candidates[51–53]. In some cases, the analysis on the field-dependent part of the specific heat shows that orphan magnetic moments trapped in defects cause a Schottky-like anomaly. Here, we have applied the same analysis on different fields. Supplementary Fig. 3 shows the difference between 0 and 2 T curves $\Delta C_{0T-2T}/T$, which can be described by putative zero-field split doublets with a level splitting $\Delta E$. A linear fit of $\Delta E$ versus $\mu_0 H$ then results in a Zeeman splitting Lande $g'$ factor. However, in our case, not only does the fit become less reliable at fields ≥4 T (large error bars as shown in the inset of Supplementary Fig. 3), but also the obtained Lande $g'$ factors are unphysically small. Therefore, we can rule out the case that the existence of defects causes the broad peak of $C_{mag}$ in zero field. To exactly understand its origin, one needs to invoke the spin model with possible exchange interaction terms, such as the dipole and octupolar interactions in the XYZ model (see the main text). Such work is beyond our scope at the current stage, but it is necessary for the future study to resolve which category of QSL $Ce_2Zr_2O_7$ belongs to.

**μSR experiments.** To determine whether the $Ce_2Zr_2O_7$ samples develop long-range magnetic order, we conducted μSR experiments at the Low Temperature Facility spectrometer and the General Purpose Spectrometer, Paul Scherrer Institut, Switzerland. At the Low Temperature Facility, we ground the sample into a coarse powder and glued it onto a silver plate using GE varnish dissolved in alcohol, which presents a negligible cross-section for muon capture. We quenched the magnet to remove stray magnetic fields while the sample was above 2 K, then cooled to base temperature and collected time spectra in zero field at temperatures from 19 mK to 750 mK. We found no evidence for long-range magnetic order, which would present as oscillations or a fast-relaxing component. Instead, the zero-field data were fitted at all temperatures with a one-component stretched exponential form, plus a simple exponential background for the ~21% of muons landing in the silver. We found evidence for a kink in the relaxation rate $1/T_1$ below 200 mK (Supplementary Fig. 4), suggesting that the dynamics begin slowing down below this temperature. To determine whether this kink is connected to the broad feature in the specific heat data, we also conducted longitudinal field measurements up to 14,000 G (1.4 T) at 19, 450 and 750 mK. At the base temperature, the muon spin decouples from the internal field near 3,000 G. At 0 and 1.4 T, the 450 mK and 750 mK curves are nearly identical and do not relax as quickly as the 8,000 and 14,000 G curves at 19 mK. This suggests that the application of an external magnetic field pushes the spin fluctuations to higher energy, as seen in the specific heat curves, and results in a slower relaxation of the muon polarization.

The overall magnitude of $1/T_1 \lesssim 0.25\,\mu s^{-1}$ in zero field suggests that if μSR is caused by static internal fields, the field strength should be on the order of $H_{int} \approx (1/T_1)/(\gamma_\mu/2\pi) \approx 50$ G, where $\gamma_\mu = 2\pi \times 135$ MHz is the gyromagnetic ratio of the muon. The experimental value determined here is much larger, suggesting that the internal fields are dynamic at the base temperature. Moreover, for all ZF and LF curves, we find that the stretching exponent is between 0.5 and 1. This corresponds to a regime with dynamically fluctuating moments that is between the dilute and dense limits, similar to μSR observations of $Ce_2Sn_2O_7$ (ref. [33]) and $SrDy_2O_4$ (ref. [54]), but different from $SrCr_8Ga_4O_{19}$ (ref. [55]).

**CEF level measurements.** Inelastic neutron scattering experiments were carried out on 5 g $Ce_2Zr_2O_7$ and 5 g $La_2Zr_2O_7$ polycrystalline samples on the fine-resolution Fermi chopper spectrometer, SEQUOIA, at the Spallation Neutron Source, ORNL. For each compound, we collected data for 4 h with 250 meV incident energy at 5, 100 and 200 K. Since $La_2Zr_2O_7$ is the non-magnetic analogue of $Ce_2Zr_2O_7$, it can serve as the background, which helps us to subtract phonon contributions from the measured spectra.

The raw data for $Ce_2Zr_2O_7$ and $La_2Zr_2O_7$, as well as the background-subtracted data set for $Ce_2Zr_2O_7$, are shown in Supplementary Fig. 5. The CEF excitations are clearly visible because the corresponding intensity decreases with $Q$ as a result of the magnetic form factor of $Ce^{3+}$. The strong atomic spin–orbit coupling of the $4f^1$ electron in the $Ce^{3+}$ ion entangles the electron spin ($S = 1/2$) with the orbital angular momentum ($L = 3$) into a $J = 5/2$ total moment. The six-fold degeneracy of the $J = 5/2$ total moment is further split into three Kramers doublets by the $D_{3d}$ crystal field. This is indeed what we observe in the neutron data: in all of the data from different temperatures, two magnetic transitions are visible at 55.8 and 110.8 meV (Supplementary Fig. 5a–c,g). These excitations are absent in the non-magnetic reference compound $La_2Zr_2O_7$ (Supplementary Fig. 5d–f). The Hamiltonian of the CEF $H_{CEF}$ in the $D_{3d}$ crystal field is $H_{CEF} = B_2^0 \hat{O}_2^0 + B_4^0 \hat{O}_4^0 + B_4^3 \hat{O}_4^3 + B_6^0 \hat{O}_6^0 + B_6^3 \hat{O}_6^3 + B_6^6 \hat{O}_6^6$, where $B_n^m$ ($n, m$ are integers and $n \geq m$) are CEF parameters that will be determined experimentally, and the Stevens operators $\hat{O}_n^m$ are polynomial functions of the components of the total angular momentum operator $J_z, J_+$ and $J_-$ ($J_\pm = J_x \pm i J_y$). The Stevens factor $\gamma = 0$ for $Ce^{3+}$ in this CEF; therefore, all three sixth-order parameters, $B_6^0, B_6^3$ and $B_6^6$, are zero. Using MantidPlot software (version 3.13), we can simultaneously fit the inverse susceptibility and inelastic neutron scattering data. The data used for the fitting are the $Q$ cut from 4 Å$^{-1}$ to 6 Å$^{-1}$ of Supplementary Fig. 5g, and the inverse magnetic susceptibility versus temperature under a magnetic field of 0.1 T. The data and the corresponding fits coincide well, as shown in Fig. 1c.

The result of the fitting is $B_2^0 = -1.27265$ meV, $B_4^0 = 0.322593$ meV and $B_4^3 = -1.85767$ meV. The eigenfunctions of the ground state are $|+3/2\rangle$ and $|-3/2\rangle$, with the eigenvalue 0 meV. The eigenfunctions of the first excited state are $0.9402|-5/2\rangle - 0.3406|+1/2\rangle$ and $0.3406|-1/2\rangle + 0.9402|+5/2\rangle$, with an eigenvalue of 55.8 meV. The eigenfunctions of the second excited state are $0.9402|-1/2\rangle - 0.3406|+5/2\rangle$ and $0.3406|-5/2\rangle + 0.9402|+1/2\rangle$, with the eigenvalue 110.8 meV.

Here we define the effective spin that operates on the ground-state dipole–octupole doublet $||\Psi^{\uparrow/\downarrow}\rangle \equiv |J^z = \pm 3/2\rangle$ for each lattice site with

$$S^x \equiv \frac{1}{2}(|\Psi^+\rangle\langle\Psi^-| + |\Psi^-\rangle\langle\Psi^+|) \quad (1)$$

$$S^y \equiv \frac{1}{2}(-i|\Psi^+\rangle\langle\Psi^-| + i|\Psi^-\rangle\langle\Psi^+|) \quad (2)$$

$$S^z \equiv \frac{1}{2}(|\Psi^+\rangle\langle\Psi^+| - |\Psi^-\rangle\langle\Psi^-|) \quad (3)$$

where $S^x$ and $S^y$ are related to the magnetic octupole moments in terms of $J$ operators, and $S^z$ is related to the magnetic dipole moment. As $S^x$ and $S^z$ transform identically under space group symmetry, $S^x$ is sometimes regarded as a dipole moment, just like the $S^z$ component. This is the symmetry point of view. However, microscopically, when $S^x$ or $S^y$ acts on the ground-state wavefunction, it is equivalent to applying $J^+$ or $J^-$ three times. This is the microscopic reason that both $S^x$ and $S^y$ are magnetic octupole moments. In the current work, we take this microscopic view and regard $S^x$ and $S^y$ as the magnetic octupole moments throughout.

**Diffuse neutron scattering.** Diffuse neutron scattering experiments were performed on a 2.3 g floating-zone-grown $Ce_2Zr_2O_7$ single crystal using the elastic diffuse scattering spectrometer, CORELLI, at the Spallation Neutron Source, ORNL[56]. The sample was aligned in the $(h, h, l)$ scattering plane on an oxygen-free copper holder and the temperature was regulated using a $^3$He inset. The scattering was performed at three different temperatures, 50 K, 2 K and 240 mK, using a white incident neutron beam. Note that the (002) Bragg peak, which is forbidden in the pyrochlore space group $Fd\bar{3}m$, can be observed at all temperatures (Supplementary Fig. 6). However, it is not due to a deviation from the pyrochlore structure, but probably due to a non-structural origin (multiple scattering), as such a peak is not seen in X-ray and neutron single crystal diffraction measurements. In addition, we did not find the (002) peak in our inelastic scattering experiments using the same sample around zero energy transfer (elastic). Note that one of the signatures of multiple scattering is that it is strongly energy dependent and occurs only at specific energies.

**Inelastic neutron scattering.** Inelastic neutron scattering experiments were carried out on a 2.3 g $Ce_2Zr_2O_7$ single crystal on the cold neutron chopper spectrometer, CNCS, at the Spallation Neutron Source, ORNL. The sample and the set-up were the same as that on CORELLI. Three incident neutron energies (1.55, 3.32 and 12.0 meV) were used with instrumental resolution at elastic positions of 0.05, 0.11 and 0.8 meV, respectively. Supplementary Fig. 7 shows raw data obtained with $E_i = 3.32$ meV at 35 mK and 12 K. The data plotted in Figs. 3 and 4 are obtained by subtracting the 12 K data from the 35 mK data. Our assumption is that the magnetic scattering at 12 K is diffusive enough and would be wavevector/energy independent, and can thus serve as the background.

### Data availability
The data that support the plots in this paper and other findings of this study are available from the corresponding author on reasonable request.